\documentclass{article}
\usepackage{bm}
\usepackage{amsmath,amssymb,graphics}
\usepackage{yfonts}
\usepackage{epsfig,latexsym}
\usepackage[colorlinks=true, pdfstartview=FitV, citecolor=red, urlcolor=blue]{hyperref}
\newcommand{\vf}{\varphi}
\newcommand{\qn}{\textswab{q}}
\newcommand{\wn}{\textswab{w}}

\newcommand{\slsh}[1]{\displaystyle{\not} #1}
\newcommand{\be}{\begin{equation}}
\newcommand{\ee}{\end{equation}}
\newcommand{\bea}{\begin{eqnarray}}
\newcommand{\eea}{\end{eqnarray}}

\begin{document}

\author{Giuseppe Policastro}
\date{}

\title{Supersymmetric hydrodynamics from the AdS/CFT correspondence }

\maketitle

\begin{center}
 \emph{Laboratoire de Physique Th\'eorique}\footnote{Unit\'e Mixte du CRNS et
    de l'Ecole Normale Sup\'erieure associ\'ee \`a l'universit\'e Pierre et
    Marie Curie 6, UMR
    8549.\\LPTENS-08/60}, 
\emph{ Ecole Normale Sup\'erieure,  \\
24 rue Lhomond, F--75231 Paris Cedex 05, France} \\
 \textrm{policast@lpt.ens.fr}
\end{center}

\vskip 1cm 

\abstract{We compute holographically the dispersion relation for a hydrodynamic 
mode of fluctuation (the phonino) of the density of supersymmetry current in ${\mathcal N}=4$ SYM at strong coupling. The mode appears as a pole at low frequency and momentum in the correlator of supercurrents. It has a wave-like propagation, and we find its speed and coefficient of attenuation. }

\newpage
\tableofcontents

\section{Introduction} 

In \cite{Policastro:2001yc, Policastro:2002se, Policastro:2002tn} the author of this note and his collaborators have proposed to look at the finite temperature phase of ${\mathcal N}=4$ SYM; general arguments lead to the prediction that the dynamics of low-lying excitations would be well described by the hydrodynamic approximation, where the only relevant degrees of freedom are the densities of conserved charges.  The conjecture was borne out by the computation 
of the correlators of conserved currents, that exhibited the expected analytic structure, namely 
a pole for small frequency and momentum corresponding to excitations with a diffusive or a wave-like behavior. The coefficients entering the dispersion relations of these excitations, called {\it transport coefficients} \cite{landau5}, capture the low-energy dynamics of the finite-temperature phase. We computed the transport properties of the stress-energy tensor and the R-currents. 
The calculation was performed holographically, using the known correpondence between 
SYM in 4 dimensions and type IIB supergravity in the $AdS_{5} \times S^{5}$ background \cite{Aharony:1999ti}. This means that the results were valid in the limit of infinite number of colors and infinite 't Hooft coupling. \\
Our calculation has subsequently been extended in very many directions, mostly trying to make 
predictions in more realistic theories by deforming the original setup, e.g. breaking conformal invariance, 
supersymmetry, adding flavors etc.  (see \cite{Kaminski:2008ai} for a partial list of relevant references). 

In this paper we fill a gap in the analysis of the linearized hydrodynamics of ${\mathcal N}=4$ SYM:  we consider the transport properties associated to the fluctuations of the supersymmetry current. 
It was first pointed out in \cite{Lebedev:1989rz} that a supersymmetric thermal medium must have a  fermionic collective excitation, that they called {\it phonino}; in that paper they discussed the case of the Wess-Zumino model, and determined the properties of the phonino for moderately small temperatures, 
but could not discuss the ultrarelativistic case, as it was too complicated. As far as we know, there have not been other attempts at computing the phonino spectrum perturbatively. 
The outline of this paper is the following: in section \ref{fieldth} we discuss the prediction for the correlators of supercurrents following from the hydrodynamical Ansatz; in section \ref{bulk} we detail the 
holographic computation of the correlators, which is essentially the solution of the classical equations of motion for the gravitino in the dual geometry. In the final section \ref{final} we summarize and briefly discuss our results.

\section{Supersymmetric hydrodynamics}\label{fieldth}

\subsection{The phonino}

We follow the treatment of \cite{Kovtun:2003vj}, to which we refer for a more extensive discussion. 
We consider a 4-dimensional supersymmetric field theory. The density of the supersymmetry charge should be included in the list of effective degrees of freedom in the long-wavelength regime, by the same arguments that hold for any conserved 
charge and imply that it will relax with arbitrarily long time scale. 
In order to turn on a vev for the density of the supercurrent we may introduce a fermionic chemical potential $\mu_\alpha, \bar\mu_{\dot\alpha}$. In the equilibrium state, the supersymmetry algebra implies \footnote{See appendix \ref{gammamatrices} for our conventions on the gamma matrices.}
\bea
\langle S_i^{\alpha}\rangle = - \beta \, \langle T_{ij} \rangle  (\gamma^{j} \bar\mu)^\alpha  \,, \nonumber\\
\langle \bar S^{\dot\alpha}_{i}  \rangle = - \beta \, \langle T_{ij} \rangle  (\gamma^{j} \mu)^{\dot\alpha} \,.
\eea 
The linearized hydrodynamics deals with small fluctuations out of equilibrium. 
We assume that the relevant degrees of freedom are contained in the density of charge $S_{t}^{\alpha} = \rho^{\alpha}$, and a constitutive relation is needed in order to express 
the other components of the current in terms of $\rho^{\alpha}$. The most general expression, 
linear in fields and derivatives, and assuming that no other vevs and chemical potentials are turned on, is the following 
\be\label{const}
 S_i^\alpha = - D_s \partial_{i} \rho^\alpha - D_\sigma (\gamma_{ij} \partial^{j} \rho)^\alpha - \frac{{\mathcal P}}{\varepsilon} (\gamma_{i}  \gamma^{1} \rho)^\alpha 
\ee
The third term in this expression is fixed by the susy algebra, and ${\mathcal P}, \varepsilon$ are the pressure and energy density in the equilibrium state. If the theory is conformal ${\mathcal P} = \varepsilon/3$; moreover $\gamma^i S_{i}$ is related by susy to $T_i^{i}$ and it has to vanish; this implies that $D_s = D_\sigma$. 
At this order, then, there is only one transport coefficient that captures the dynamics of the fermionic fluctuations. 
The conservation of the current, $\partial_t \rho + \partial_j S^j=0$, closes the system of equations and allows one to find the existence of waves propagating in the 
medium; they are solutions of the form 
\be\label{phonino}
\rho_\alpha(t) = e^{-D_s k^2 t} \left( \delta_\alpha^\beta cos(k v_s t) + ( k_i \cdot \gamma^i \gamma^{1} ) sin(k v_s t) \right) \, \rho(0) 
\ee
with a velocity $v_s = \frac{{\mathcal P}}{\varepsilon}$, to be compared with the usual speed of sound $c_{s}^2 = \frac{\partial {\mathcal P}}{\partial\varepsilon}$. The transport coefficient $D_s$  determines the 
damping of this excitation, that has been called ``phonino'' . 

In presence of a supercharge density, other conserved currents can also be affected via non-linear terms in their constitutive relations. The R-current receives a contribution 
$$J_{R}^{i} = \ldots - {{\mathcal P}\over \varepsilon^{2}} \, \bar \rho \gamma^{i} \rho $$
(where the dots are for terms not depending on $\rho$) 
and the coefficient $D_{s}$ enters in the R-current correlators; in \cite{Kovtun:2003vj} it is shown that such non-linear and non-derivative terms in the constitutive relations make the correlator decay more slowly in time, with a power law instead of the exponential decay characteristic of diffusion processes. 
The leading large-time behavior of R-current correlators was found to be 
\be
\langle J_{R}^{k}(t) J_{R}^{l}(0) \rangle \sim {\delta^{kl} \over 12 (\pi t)^{3/2}} \left( {T \over \bar w} {\chi_{R} \over (D_{R} + \gamma_{\eta})^{3/2}} + {1\over 4}{c_{s}^{2} \over (2 D_{s})^{3/2}} \right) 
\ee
where $\bar w = \varepsilon+{\mathcal P}$, $\gamma_{\eta} = \eta/\bar w$, $\chi_{R}$ and $D_{R}$ are the charge susceptibility and diffusion constant. All these quantities have been computed for ${\mathcal N}=4$ SYM, with the exception (up to now) of $D_{s}$.

\subsection{Supercurrent correlators }

The presence of the fluctuation mode (\ref{phonino}) has to be reflected in the 
structure of the correlators of operators with the same quantum numbers (this is the 
statement of the fluctuation-dissipation theorem). 
In particular we are interested in the retarded correlator of the supercurrents $S_{i}$, defined as 
\begin{equation}\label{supercorr}
G_{ij}^{\dot\alpha \beta} (k) = \int d^{4} x e^{-i k\cdot x} \theta(x^{0}) \langle \{  \bar S_{i}^{\dot \alpha} (x), \, S_{j}^{\beta}(0) \} \rangle \,. 
\end{equation}
The correlator (\ref{supercorr}) is constrained by the Ward identities: 
\be
k^{i} G_{ij} = 0 \, , \quad \gamma^{i} G_{i j} =  G_{i j} \, \gamma^{j} = 0 \,.
\ee
The first equation is simply the conservation of the current, while the second follows from 
 superconformal invariance as already mentioned. The correlator can be built out of projectors, that automatically implement the identities. 
First we analyse the case of zero temperature, where the correlator has to respect the 4-dimensional 
Lorentz invariance. The projector on the transverse gamma-traceless part of a spinor-vector is 
\begin{equation}\label{proj}
P_{i}^{~j} = \delta_{i}^{j} - \frac{1}{3} \left( \gamma_{i} - \frac{k_{i} \slsh{k}}{k^{2}} \right) \, \gamma^{j}
 - \frac{1}{3 k^{2}} \left(4 k_{i} - \gamma_{i} \slsh{k} \right) \, k^{j} \,.
 \end{equation}
The correlator can then be written as $G_{ij} = P_{i}^{~k} M_{kl} P^{l}_{~j}$. The matrix $M$ can be expanded on a basis of gamma matrices, and the only form allowed by the symmetries is 
$$M_{kl} = A(k^{2}) \, \slsh{k} \, \eta_{kl} \,.$$ 
The zero-temperature correlator is then completely fixed up to a scalar function of momenta, $A(k^{2})$.
 
At finite temperature, only spatial rotational invariance is preserved, and the correlator can depend, apart from $k$, also on another vector that is the velocity of the fluid; since it is at rest, this is $u^{i} = (1,0,0,0)$. Let us write $k^{i} = (\omega, {\bf q})$.

We can form another  projector with the properties of (\ref{proj}), that is also transverse to $u^{i}$: 
\bea
P^{T}_{1 1} = && P^{T}_{1 i} = 0 \, \nonumber\\
P^{T}_{jk} = && \delta_{jk} - {1\over 2} (\gamma_{j} - \frac{q_{j} \slsh{q}}{q^{2}}) \gamma_{k} - {1\over 2 q^{2}} (3 q_{j} - \gamma_{j} \slsh{q}) q_{k} 
\eea
 and define $P^{L} = P - P^{T}$ so that $P^{L} P^{T} = 0$. The correlator can have several possible structures: $P^{L} P^{L}, P^{T} P^{T}$, and $P^{L} P^{T} + P^{T} P^{L}$. We are interested in the 
 fluctuations of the density, which are contained in the longitudinal part, so we can write  
\be\label{generalcorr1}
 G_{ij} = P^{L}_{ik} M_{km} P^{L}_{mj}
 \ee
 and now the form of $M$ allowed by the symmetries is 
 \begin{eqnarray}\label{generalcorr2}
 M_{km} =  \slsh{k} \, C_{km} +  \slsh{u} \, C'_{km} \,, \nonumber\\
 C_{km} = a \, (\eta_{km} + b \, u_{k} u_{m}) \,, \\
 C'_{km} = a' (\eta_{km} + b' u_{k} u_{m}) \,. \nonumber
 \end{eqnarray}
 Not all the functions $a,b,a',b'$ are independent, it turns out that the following combinations vanish 
 inside the projectors: 
 \begin{eqnarray*}
 \slsh{k} \, (\eta_{km} + {3 k^{2} \over 2 {\bf q}^{2}} \, u_{k} u_{m} )\,, \\
 - \slsh{k} {\omega \over {\bf q}^{2} } u_{k} u_{m} + \slsh{u} \, ( \eta_{km} + {k^{2} \over 2 {\bf q}^{2}} u_{k} u_{m} ) \,.
 \end{eqnarray*}
 Thus for generic values of the momenta we can fix two of the parameters. We haven't been able to find a particularly nice canonical choice of structures. 
 The correlator is then determined, in this channel, by 2 functions of ${\bf q},\omega$. The hydrodynamical prediction is that 
these functions  should have a pole, in the complex $\omega$ plane, that goes to the origin as ${\bf q} \to 0$.

\section{Holographic derivation} \label{bulk}

\subsection{Zero temperature}

The two-point correlator of the supersymmetry current in ${\mathcal N}=4$ SYM at zero temperature 
has been computed in \cite{Corley:1998qg, Volovich:1998tj}. We recall in this section the main points of the derivation, to be used for reference in the next sections. 

The operator dual to the susy current is the gravitino, a Rarita-Schwinger field propagating in the dual 
$AdS$ geometry. We follow here the presentation of \cite{Rashkov:1999ji}. 
The bulk action is 

\begin{equation}
 S = \int d^{d+1}x \sqrt{g} \left( \bar\Psi_\mu \Gamma^{\mu\nu\rho} D_\nu \Psi_\rho - m \bar\Psi_\mu \Gamma^{\mu\nu} \Psi_\nu \right) \,, 
\end{equation}  
where $\Gamma^{\mu} = e^{\mu}_{a} \gamma^{a}$. 
For the time being we can leave $d$ and $m$ unspecified. 
The action is evaluated on the (Euclidean) $AdS_{d+1}$ background with metric 
\begin{equation} 
 ds^2 = \frac{1}{x_0^2} (dx_0^2 + \delta^{ij} dx_i dx_j) \,.
\end{equation}
The study of the solutions of the equations of motion in the neighborhood of the boundary 
$x_0 = 0$ shows that components of different chirality have different asymptotic behavior. In terms of 
the spinor $\psi_a = e_a^{~\mu} \Psi_\mu$ one has 
\begin{eqnarray}
\psi_i^+(x_0,x_i) &=& (x_0)^{d/2 -m} \varphi_i({\bf x}) + \ldots \,, \nonumber \\
\psi_i^-(x_0,{\bf x}) &=& (x_0)^{d/2+m} \chi_i({\bf x}) + \ldots 
\end{eqnarray}

The solution is required to be regular in the bulk. It is possible to show that this 
imposes a relation between $\varphi_i$ and $\chi_i$, 
so that only one of them can be assigned as a boundary condition. Without loss of generality one can take $m>0$, then the appropriate boundary spinor can be shown to be $\varphi$. Moreover, it has to 
satisfy a constrain $\gamma^i \, \varphi_i = 0 $. 
The complete solution can then be expressed in terms of the boundary data :
\be\label{volovich}
\psi_i (x_0,{\bf k}) =
x_0^{d/2-m} \left[ \varphi_i
-\frac{i\slsh{k}}{k} \frac{K_{m-1/2}}{K_{m+1/2}}\varphi_i
- \right. 
\ee
$$
\left( i x_0 k_i \frac{K_{m-1/2}K_{m+3/2}-K^2_{m+1/2}}
{K_{m+1/2}}
 +(\frac{i\slsh{k}\gamma_i}{k})K_{m-1/2}\right) \times
$$
$$
\left.  \frac{2k_j \slsh{k}}{k^2} 
\frac{1}{(2m+d-1)K_{m+1/2}-2x_0  k K_{m+3/2}}
\varphi_j  \right]. $$
Here the Bessel functions have argument $K_\nu (k x_0 )$. A similar expression holds for 
the conjugate field $\bar\psi$. 

We recall here one of the basic tenets of the AdS/CFT correspondence: the on-shell action, with given asymptotic conditions for the fields, is the generating functional for the correlators of CFT operators. The bulk action vanishes on-shell for fermionic fields, therefore the only contribution comes from  the boundary action \cite{Henningson:1998cd}
\be 
S_{bdy} = \int d^d {\bf x} \sqrt{h} h^{ij} \bar\Psi_i  \Psi_j \,.
\ee 
Here $h_{ij}$ is the induced metric on the boundary.  
The action has to be regularized by evaluating the integral on a surface $x_0 = \epsilon$, and taking the limit $\epsilon \to 0$. In the limit, the expression contains divergencies that in principle should be subtracted by introducing suitable counterterms, however one can see that the divergent pieces are analytic in $k$, therefore they only affect the correlators by contact terms. Finally, the regularized action can be written as 
\be\label{bdyaction}
S_{bdy} = \int d{\bf k} \, \delta^{ij} \left( \bar \varphi_i (-k) \chi_j(k) + \bar\chi_i(-k) \varphi_j(k) \right)
\ee
the boundary value $\varphi$ is the source for the field theory operator, and the appearance of a non-trivial correlator is due to the 
nonlocal relation that expresses $\chi$ in terms of $\varphi$, as implicitly given in (\ref{volovich}). 
The result is, up to an overall factor,  
\be 
G_{ij} = \Pi_i^r \, \frac{\slsh{k}}{k} \, ( \delta_{rs} - \frac{2(2 m + 1)}{(d+2m+1)} \frac{k_r k_s}{k^2} ) \, \Pi^s_j \,,
\ee
where $\Pi_i^r = \delta_i^r - \frac{1}{d} \gamma_i \gamma^r $ is the projector on the gamma-traceless part, since the source is gamma-traceless. The correlator is transverse, $k^i G_{ij} = 0$, for $m=\frac{d-1}{2}$, that is when the gravitino is massless, as it should be. In the case of interest, $d=4$ and $m=3/2$, in units of the AdS radius.

\subsection{Finite temperature}

\subsubsection{The gravitino in the near-extremal 3-brane background}

In order to compute the correlator at finite-temperature, we need to study the Rarita-Schwinger equation in the background of a non-extremal black 3-brane. In this case we are not able to solve the equations for generic mass and dimension, so we work directly in $d=4$ and $m=3/2$. 
The near-horizon geometry has the following metric : 

\begin{equation}\label{metric}
ds^2 = {\pi^2 T^2 R^2\over u} \left( - f(u) d t^2 + 
d x^2 + d y^2 + d z^2\right) +{R^2 \over 4 f(u) u^2} d u^2
\end{equation}
where $f(u) = 1-u^{2}$. \\
The radial coordinate $u \in [0,1]$; the boundary is at $u=0$ and there is a horizon at $u=1$. Near the 
boundary, $u$ is related to $x_0$ of the previous section by $u=x_0^2$. 
The covariant derivative on a spinor, in this background,  is 
\begin{equation}
D_{\mu} = \partial_{\mu} + {1\over 4} \omega_{\mu}^{ab} \gamma_{ab}\,,
\end{equation}
i.e.
\begin{eqnarray}
D_{u} &=& \partial_{u}\,, \nonumber\\
D_{i}  &= & \partial_{i} + {1\over 2} \omega_{i}^{5 (i)} \gamma_{5 (i)}\,,
\end{eqnarray}
where $i=t,x,y,z$ and $(i)=1,2,3,4$. The non-vanishing components of the spin connection are 
\begin{eqnarray*} \label{SP1}
\omega^{5 1}_t &=&   {\pi T (1+u^2)\over \sqrt{u}} \\
 \omega^{5 2}_x =  \omega^{5 3}_y =  \omega^{5 4}_z&=& \pi T \sqrt{{f\over u}}\\
\end{eqnarray*}
The metric ( \ref{metric} ) satisfies $R_{\mu\nu} = \Lambda g_{\mu\nu}$ with $\Lambda = -4/R^{2}$. 

The Rarita-Schwinger equation is:  

\be\label{RSE}
\Gamma^{\mu\rho\sigma} D_{\rho} \Psi_{\sigma} - m \Gamma^{\mu\nu} \Psi_{\nu} = 0 \, .
\ee
The equation is invariant under a gauge transformation parametrised by an arbitrary spinor $\epsilon$:  
\be
\delta \Psi_{\mu} = D_{\mu} \epsilon + \frac{1}{3} m \Gamma_{\mu} \epsilon \,. 
\ee
We show in appendix \ref{GT} that the RS equations imply, in $d$ space-time dimensions, 
\be\label{RSconstr}
\left( {2 - d \over 4} \Lambda - m^{2} {d-1 \over d-2} \right) \Gamma \cdot \Psi = 0 \,.
\ee
For a generic value of the mass, (\ref{RSconstr}) implies $\Gamma\cdot\Psi=0$, but for $d=5$, 
$\Lambda = - 4/R^{2}$  and $mR = 3/2$, the coefficient in the brackets vanishes, so the constraint is not implied 
by the equations of motion. However, it can still be imposed as a gauge-fixing condition. We will work in this gauge.  The relation 
(\ref{RSE2}) then implies that $D \cdot \Psi = 0$.  In holographic computations it would be perhaps more natural to impose an ``axial'' gauge $\Psi_{u}=0$, but we were not able to solve the equations of motion in that gauge; at the level of two-point functions it does not make a difference, but it could be an issue in the computation of higher-order correlators. 

The equations of motion in this gauge can be written as 
\begin{equation}
\Gamma^{\mu} \left( D_{\mu}\Psi_{\nu} - D_{\nu}\Psi_{\mu}\right) + 
m \Psi_{\nu} = 0 
\label{rarita}
\end{equation}
Explicitly, and again redefining the field as $\psi_a = e_a^\mu \Psi_\mu$, eqs. (\ref{rarita}) read ($\psi' = \partial / \partial_u \psi$): 

\begin{eqnarray}\label{liliaequ}
&\,&  \gamma^5 \psi_5 ' + {1\over 2 \pi T \sqrt{u f}} \left( {1\over \sqrt{f}} \gamma^1 
\partial_t  + 
\gamma^{j} \partial_{j} \right) \psi_5
\nonumber \\ 
& + &  {2 u^2-3\over 2 u f} \gamma^5\psi_5 
 +  {u\over f} \gamma^1\psi_1 + {m R\over 2 u \sqrt{f}} \psi_5 =0 
\end{eqnarray}


\begin{eqnarray}\label{liliaeqt}
&\,&  \gamma^5 \psi_1 ' + {1\over 2 \pi T \sqrt{ u f}}
\left( {1\over \sqrt{f}} \gamma^1 \partial_t  + 
 \gamma^{j} \partial_{j} \right) \psi_1  \nonumber \\ &+& 
{ u^2-2\over 2 u f}\gamma^5 \psi_1 + {1+u^2\over 2 u f} \gamma^1 \psi_5 +
{m R\over 2 u \sqrt{f}} \psi_1 =0
\end{eqnarray}


\begin{eqnarray}\label{liliaeqj}
&\,&  \gamma^5 \psi_k ' + {1\over 2 \pi T \sqrt{ u f}}\left( {1\over \sqrt{f}} \gamma^1 \partial_t  + 
\gamma^{j} \partial_{j} \right) \psi_k  \nonumber \\ &+& 
{ u^2-2\over 2 u f}\gamma^5 \psi_k - {1\over 2 u} \gamma_{k} \psi_5 +
{m R\over 2 u \sqrt{f}} \psi_k =0 \, 
\end{eqnarray}
$$  j,k=x,y,z $$
Even though we have fixed a gauge, there is still a residual gauge symmetry parametrised by $\epsilon$ such that $\slsh{D} \epsilon + { 5 m \over 3} \epsilon =0$, or 
\be\label{residgauge}
\gamma^{5} \epsilon' + {1\over 2 \pi T \sqrt{uf}} \left( {1 \over  \sqrt{f}} \gamma^{1} \partial_{t} + \gamma^{j} \partial_{j} \right) \epsilon + \left( {u^{2}-2 \over 2 u f}
\gamma^{5} + {5 m R \over 6 u \sqrt{f}} \right) \epsilon = 0 \,.
\ee
Notice that this equation determines only the radial profile of $\epsilon$, leaving arbitrary dependence on the spacetime coordinates.

\subsubsection{Solution of the equations of motion}

The equations (\ref{liliaequ},\ref{liliaeqt},\ref{liliaeqj}) are a complicated system that cannot be solved exactly. 
We want to solve them  in perturbation theory in the momentum. We consider a fluctuation with a definite momentum $k^\mu = 2 \pi T \, (\wn,0,0,\qn)$. 

It is useful to decompose the fluctuations of the fields according to their spin under the 
unbroken transverse $O(2)$ (rotations in the $xy$ plane). There is one component of spin 3/2, given by $\eta \equiv \gamma^{2} \psi_{2} - \gamma^{3} \psi_{3}$, and four components of spin 1/2, given by $\psi_{1}$, $\psi_{4}$, $\psi_{5}$ and $\phi \equiv \gamma^{2} \psi_{2} +\gamma^{3} \psi_{3}$. Components of different spin are decoupled, and we are interested in the sector of spin 1/2, since it contains the time component of the gravitino, which is dual to the supercharge density on the boundary, and we expect the correlators of the charge density to exhibit the hydrodynamic behavior. In this sector we can then set $\eta=0$. 
Moreover, we can use the matrix $\gamma^{23}$ as a chirality matrix; it commutes with the evolution, as one can see by inspection of the e.o.m., therefore we can choose to study the part of definite $\gamma^{23}$ chirality; the number of components is then reduced by half. 

A further simplification comes from the fact that the gauge condition $\gamma^{5} \psi_{5} + \gamma^{i} \psi_{i} =0$, together with the equations of motion, yields the 
following condition (compare with (2.17) of \cite{Corley:1998qg} ): 

\begin{equation}\label{algrel}
\left(4 \gamma_{5} \slsh{P} + {u^{2} -3 \over  \sqrt{u f}} - {3 \over  \sqrt{u}} \gamma_{5} \right) \gamma^i \psi_i + \left( {2 u^{3/2} \over  \sqrt{f}} \, \gamma^{1} - {4 i \wn \over \sqrt{f}} \, \gamma_{5} \right) \, \psi_{1} -4  i \qn \gamma_{5}\, \psi_{4} = 0 
\end{equation}
where $\slsh{P} = -i \wn/\sqrt{f} \gamma^{1} + i \qn \gamma^{4}$.

We can use the relation (\ref{algrel}) to solve for $\gamma\cdot\psi \equiv \gamma^i \psi_i$ in terms of $\psi_{1}$ and $\psi_{4}$. Then we derive a system of equations for the two latter fields. Again we consider the $\gamma^{23}$-chiral part. Finally, we have a system of 4 scalar equations for the remaining components that we write as a 4-vector $X = (\psi_1^+, \psi_1^-, \psi_4^+, \psi_4^-)$ (here the superscript refers to the usual chirality in 4d).

The resulting system has regular singular points at the boundary and at the horizon. 
%
The local solutions at the horizon are 

\begin{eqnarray*}
(1-u)^{-\frac34-i \frac\wn2} \, \left(  i , 1 , 0 , 0 \right) \\
(1-u)^{-\frac14-i \frac\wn2} \, \left(  0 , 
       0 , i , 1  \right)  \\
(1-u)^{-\frac34+ i \frac\wn2} \left( -i ,1,0,0  \right)  \\
(1-u)^{-\frac14+ i \frac\wn2} \left(  0 , 0 , -i ,1  \right)
\end{eqnarray*}
The first two eigenvectors corrispond to the incoming-wave boundary condition. As the previous finite-temperature computations have shown \cite{Policastro:2002se}, one has to impose incoming-wave boundary condition at the horizon in order to recover the retarded correlator.  The form of the eigenvalues shows that we can eliminate the outgoing wave solutions by imposing the boundary condition $(1+ i \gamma_{5} )\psi_{1,4} = 0$, or $\psi^{+} = i \psi^{-}$. 

The solution at order zero with these boundary conditions is 

\begin{equation}\label{chi0}
X_{(0)}  = f^{-3/4}
\left(
\begin{array}{l}
 i  \, u^{9/4} \left(2-\sqrt{f}\right) (1+\sqrt{f})^{-1/4} \alpha _0 \\
 (1+\sqrt{f})^{1/4} \left(2+\sqrt{f}\right) u^{7/4}  \alpha_{0}\\
 i u^{1/4}\sqrt{f} \left(1+\sqrt{f}\right)^{3/4}  \left(\sqrt{f} \alpha _0+i \beta _0\right) \\
 i u^{7/4} \sqrt{f}\left(1+\sqrt{f}\right)^{-3/4} \left(i \sqrt{f} \alpha _0+\beta _0\right)
\end{array}
\right)\end{equation}
The solution depends on two parameters, $\alpha_0$ and $\beta_0$. It can be checked that under the action of the residual gauge transformation (\ref{residual}), $$\delta \alpha_0 = i a_+ ,\,  \delta \beta_0 = - a_+ .$$ 
The combination $\alpha_0 + i \beta_0$ is gauge invariant. We could use this freedom to fix one of the two parameters to zero, however we find it convenient not to do so and to keep the gauge invariance.  
At the next order, we have an explicit solution $X_{(1)}$, whose form is very complicated, and is written in appendix \ref{solfirst}. The important point is that in solving the equations in perturbation theory, at every stage there are 4 new integration constants to be determined. Again two of them can be fixed by imposing the incoming-wave condition at the horizon. A natural choice would be to require that each component has a prescribed behavior near the horizon, namely 
$$\psi \sim f^{-i \wn/2} (1+ F(u)) $$ 
with $F$ a function of order 1 in momenta, and such that $F(u=1) = 0$. It turns out that this requirement is incompatible with the gauge symmetry. However, the two remaining integration constants can be reabsorbed into a redefinition of $\alpha_0$ and $\beta_0$ to first order in momenta. It is not necessary then to keep track of them explicitly if we do not assume that $\alpha, \beta$ are homogeneous functions of the momenta.  The following non-obvious change of variables is convenient 
\be
\alpha = \frac{\alpha_0 + i \beta_0}{2} \,, \quad \beta = \frac{(\qn - 6 \wn) \alpha_0 -i \qn \beta_0}{2} \,.
\ee
The boundary value of the spinors is given in terms of these parameters by 
\be 
\left( \begin{array}{c}
 \psi_1^+ \\ \psi_4^+ 
\end{array} \right) \sim 2 (2 u)^{1/4} \, \left( 
\begin{array}{c}
    \beta  \\
  (i \sqrt{2} + \frac{10}{3} \qn + 2 \wn (5 - \sqrt{2} \L + \frac{12 \wn}{\qn - 3 \wn})) \alpha + \frac{3 \qn - \wn}{\qn-3\wn} \beta   
\end{array}
\right)
\ee
where $\L = \log \left(1+\sqrt{2}\right)$.
Solving for $\alpha, \beta$ in terms of $\varphi_1, \varphi_4$ yields 
\be
\left( \begin{array}{c}
 \alpha \\ \beta  
\end{array} \right) = 2^{-5/4} 
\left( \begin{array}{c} 
\frac{( 3 \qn - \wn) \vf_1 - (\qn - 3 \wn) \vf_4}{P(\qn,\wn)} \\ 
\vf_1
\end{array} \right)
\ee
where $$P(\qn,\wn) = - \sqrt{2} i (\qn-3\wn) -\frac{10}{3} \qn^2 + 6 \wn^{2}+ 2 \sqrt{2} \L \wn (\qn  - 3   \wn)  \,.$$
This polynomial has a zero for small momenta  
at $$\wn = \frac{\qn}{3} - \frac{4 \sqrt{2} i}{9} \, \qn^2 + {\cal O}(\qn^4) \,.$$
This pole corresponds to the propagation of a mode with speed $v= \frac13$ and an attenuation given by the 
imaginary part of the dispersion relation. The value we find corresponds to a diffusion constant 
\be\label{diffcost}
2 \pi T D_{s} = \frac{4 \sqrt{2}}{9} 
\ee
This is our main result. 

The reader may have noticed that while $\alpha$ was supposed to be a gauge-invariant quantity, its solution in terms of the boundary value is not easily recognized as such. However, just as in the zero-temperature case, the boundary value of the spinor is traceless, and the residual gauge transformation preserves this property. 
From (\ref{residgauge}) we find that the asymptotic boundary behavior of the gauge parameter is 
\begin{eqnarray}\label{residual}
\epsilon_{+} & = & a_{+} u^{-1/4} - {1 \over 3} i \slsh{k} a_{-} u^{11/4} \,, \nonumber \\
\epsilon_{-} & = & a_{-} u^{9/4} - {1 \over 2} i \slsh{k} a_{+} u^{1/4} \,, \quad \slsh{k} = - \wn \gamma^{1} + \qn \gamma^{4}
\end{eqnarray}
The transformation (\ref{residual}) acts on the boundary values as 
\bea 
\delta \psi_i &=& ( k_i -\frac14 \, \slsh{k} \gamma_i) a_+  \nonumber \\
\delta \vf_1 & \propto & (\qn - 3 \wn) a_+ \\
\delta \vf_4 & \propto & (3 \qn - \wn) a_+  \nonumber 
\eea
and the combination appearing in $\alpha$ is invariant under this transformation. 

In order to compute the correlator we need to find the negative-chirality component of the fields at the boundary. The asymptotic behavior of the negative chirality part is 
$$\psi^- \sim u^{3/4} \tilde \chi + u^{7/4} \chi $$
and since in the boundary action the determinant of the induced metric gives
a factor of $1/u^2$, and $\vf \sim u^{1/4}$, one can see that $\tilde\chi$ contributes to the divergent part of the action, that has to be subtracted in the regularization. The subleading term $\chi$ is the one that contributes to the finite part of the action. It is given by 
\be \label{chi}
2^{-1/4} \left( \begin{array}{c}
 \chi_1 \\ \chi_4
\end{array} \right) = 
\frac{3  \alpha}{\qn - 3 \wn}  
\left( \begin{array}{r}
 \qn  \\ -  \wn  \end{array} \right)  \, + 
\frac{\beta}{\qn - 3 \wn}  \left( \begin{array}{r}
3  \\ - 1  \end{array} \right) 
\ee
The other components  $\chi_{2}, \chi_{3}$ are determined by the algebraic constraints 
$\gamma^i \chi_i = 0$ and $\gamma^2 \chi_2 - \gamma^3 \chi_3 = 0$.

We can write the solution as a sum of two terms that depend on $\alpha$ and $\beta$ respectively. 
Let us consider the $\alpha$-dependent part: it depends on $\vf_{i}$ only through the gauge-invariant quantity $\alpha$, and it is transverse: $k^{i} \chi_{i} =0$; since the source is gamma-traceless, it is clear that 
inserting this term in the action (\ref{bdyaction})  yields a correlator with the right properties, 
namely a conserved and gamma-traceless one. 
We can also put the result in a covariant form. We need to repeat the analysis for the components of the fields with the opposite $\gamma^{23}$ chirality. The equations are the same after the exchange  $\psi_4 \to - \psi_4, \qn \to -\qn$. 
At the end we find an expression for the correlator that matches the general form (\ref{generalcorr1},\ref{generalcorr2}), with 
\bea
& a = 0, &\quad b' = {2 \over 3} \\
 & a' = &{81 \over 2\sqrt{2}} {(\qn^{2}-\wn^{2})^{2} \over (\qn^{2} - 9 \wn^{2})^{2}} \,. \nonumber
\eea

The second term in (\ref{chi}), depending on $\beta$, is more troubling since it is gauge-dependent. However it does not contain the hydrodynamic pole, and under a gauge transformation the denominator is cancelled; this means that the contribution of $\chi_\beta$ generates a correlator that satisfies the Ward identities up to contact terms. 
In general we would expect such contact terms, on the basis of the following argument: the susy transformations in the bulk do not act only on the gravitino, but also on the metric (and on the gauge field, although this is not relevant for us). The susy variation of the vielbein is $\delta e_{\mu}^{a} = \bar \epsilon \gamma^{a} \Psi_{\mu}$. When we try to prove the Ward identities, we should also include the boundary terms that depend on the metric; the boundary action is then : $S_{bdy} = S_{0} + S_{2}$, 
where $S_{0}$ contains  the Gibbons-Hawking term and the boundary volume, and $S_{2}$ is the term quadratic in the fermions that we have been using so far. Then under a susy transformation 
we have 
$$0 = \frac{\delta S}{\delta \Psi_{i} } (D_{i} \epsilon + \frac{m}{3} \Gamma_{i} \epsilon) + 
\frac{\delta S}{\delta e_{i}^{a}} \,  \bar \epsilon \gamma^{a} \Psi_{i} \,.$$
Taking the derivative with respect to $\bar \Psi$, and then setting $\Psi$ to zero, we find the 
modified Ward identity 
\begin{equation}  \langle \partial_{i} S^{i} \, \bar S^{j} \rangle + \frac{m}{3}  \langle (\gamma\cdot S) \bar S^{j } \rangle + (\gamma_{i})\, \langle T^{i j } \rangle = 0 
\end{equation}
However the precise form of the contact terms we find does not match the expectations, since $\beta$ only depends on $\vf_1$ whereas the expected form, as appears in the formula above,  looks like $\langle T^{i j} \rangle \gamma_i \vf_j$ and so it depends also on $\vf_4$. It is possible that 
one should add some additional counterterms to the regularized action in order to 
restore the correct Ward identities \cite{Skenderis:2002wp}, but we will not attempt to find them in the present paper. Nevertheless, we can still extract some information from the $\beta$-terms. 

Given the structure of the boundary action, one sees that $\chi$ as a function of $\vf$ can be directly interpreted as the expectation value of the supercurrent in the presence of sources. In the hydrodynamic regime, they should then 
satisfy the constitutive relations (\ref{const}). It is possible to check this on our solution, in an expansion in momenta. Notice first that $\chi_\beta$ is of order -1, whereas $\alpha$ and $\chi_\alpha$ start from order 0.  
Then the constituent relation at the lowest order, that is -1 in our case, reads 
$$ S^4_{(-1)} = \frac{1}{3} S^1_{(-1)} $$ 
and it is satisfied by the second term of (\ref{chi}). 
At the next order, from (\ref{const}) we derive 
$$S^4_{(0)} = \frac{1}{3} S^1_{(0)}  +  i D_{s} \, (2 \pi T \qn) \, S^1_{(-1)} $$ 
and one can check that this relation is satisfied, again only up to contact terms, 
with $D_{s}$ given by the expression in (\ref{diffcost}). 
This provides an alternative and more direct derivation of the value of the diffusion constant, more similar in spirit to the approach of \cite{Bhattacharyya:2008jc}, though of course at the linearized level the two methods are equivalent. 

\section{Summary and discussion}\label{final}

We have shown that the holographic description of a supersymmetric gauge theory plasma allows to exhibit the presence of a collective fermionic excitations, the phonino, 
whose existence can be predicted on general grounds; this excitation propagates like a sound wave, with a characteristic dispersion relation; the holographic method allows us to compute the lowest order terms in the dispersion relation, related to the speed and the attenuation. The results are valid in the limit of strong coupling.  
We find a finite value for the supercharge diffusion constant in this limit. The same was true for the other transport coefficients, like the shear viscosity and the R-charge diffusion constants, whose perturbative estimates indicated a vanishing value at infinite coupling. To our knowledge a perturbative calculation of $D_s$ in a gauge theory has not been done. 
In this paper we have studied only the longitudinal part of the correlators, that involves the density fluctuations. The transverse part could also be considered, and although it should not contain propagating modes, it should be possible to extract the value of the diffusion constant via a Kubo formula (see the footnote 32 in \cite{Kovtun:2003vj}). This would provide yet another confirmation of 
our result.  \\
It is tempting to speculate that $D_s$ may have a universality property analogous to the shear viscosity, namely it has the same value in all supersymmetric strongly coupled theories with a dual gravity description. It should not be too difficult to 
prove this using the methods of \cite{Bhattacharyya:2008jc} or in the membrane paradigm (see 
 \cite{Iqbal:2008by}).

\section*{Acknowledgments} 

I owe a great debt to Andrei Starinets, who collaborated in the early stages of the project. 
I thank C. Bachas, J. Troost and K. Zarembo for useful discussions, L. Yaffe for pointing out 
an imprecision, and the Perimeter Institute where 
part of the work was done. The work was supported partly by the EU under the contract MRTN-CT-2004-005104 and PIEF-GA-2008-221026. 

\appendix

\section{Gamma matrices}\label{gammamatrices}

We will use the following explicit representation of the five-dimensional flat gamma-matrices as 2x2-block matrices :

\begin{eqnarray} \label{gamma}
\gamma^{1} &=& i \left( \begin{array}{cc} 0 & 1 \\ 1 & 0 \end{array}\right) \qquad \gamma^{k+1} = i \left( \begin{array}{cc} 0 & \sigma^k \\ -\sigma^k & 0 \end{array}\right) \,\, , k = 1,2,3\nonumber\\
 \gamma_5 &=& \left( \begin{array}{cc} 1 & 0 \\ 0 & -1 \end{array}\right) \, ,
\end{eqnarray}
where $\sigma^k$ are the usual Pauli matrices. They satisfy $\{\gamma_i, \gamma_j \} = 2 \eta_{ij}$ with $\eta_{ij} = diag (-1,1,1,1,1)$. Notice that we call 1 the time component. \\
In the text we often write a gamma matrix acting on a chiral spinor, it is then understood that the matrix should be replaced by the corresponding block with the right chirality: $(\gamma^{i} \psi)_{\alpha} = (\sigma^{i})_{\alpha \dot\beta} \psi^{\dot\beta}$. Chiral indices can be raised and lowered with $\epsilon_{\alpha \beta}, \epsilon_{\dot\alpha \dot\beta}$.

 \section{Gamma-tracelessness}\label{GT}

We follow the computation in \cite{Grassi:2000dm}: Using the identity $\Gamma^{\mu\rho\sigma} = \Gamma^{\mu}\Gamma^{\rho} \Gamma^{\sigma} - g^{\mu\rho} \Gamma^{\sigma} - g^{\rho\sigma} \Gamma^{\mu} + g^{\mu\sigma} \Gamma^{\rho}$, and that $[D_{\mu},\Gamma_{\nu}]=0$, we find 
\be\label{RSE1}
\Gamma^{\mu} \slsh{D} (\Gamma \cdot \Psi) - D^{\mu} (\Gamma \cdot \Psi) - \Gamma^{\mu} D \cdot \Psi +
 \slsh{D} \Psi^{\mu} - m \,  \Gamma^{\mu\nu} \Psi_{\nu} = 0 \,.
 \ee
Contracting the last equation with $\Gamma_{\mu}$ we find the relation 
\be\label{RSE2}
\slsh{D} (\Gamma \cdot \Psi ) - D \cdot \Psi = {d-1 \over d-2} \, m \Gamma \cdot \Psi \,.
\ee
This can be plugged back in (\ref{RSE1}) to give, in $d=5$, 
\be\label{rarschw}
\Gamma^{\nu} (D_{\nu} \Psi_{\mu} - D_{\mu} \Psi_{\nu}) + {m \over 2} \, \Gamma^{\nu} \Gamma_{\mu} \Psi_{\nu} + {5 m \over 6} \, \Gamma_{\mu} \Gamma^{\nu} \Psi_{\nu} = 0 \,.
\ee
Applying $D_{\mu}$ to (\ref{RSE1}) we have 
\be\label{DRSE1}
\slsh{D} \slsh{D} (\Gamma \cdot \Psi) - D^{2} (\Gamma \cdot \Psi) - \slsh{D} D \cdot \Psi + D_{\mu} \slsh{D} 
\Psi^{\mu} - m \Gamma^{\mu\nu} D_{\mu} \Psi_{\nu} = 0 \,.
\ee
Using $\slsh{D} \slsh{D} - D^{2} = {1 \over 2} \Gamma^{\mu\nu} [D_{\mu},D_{\nu}]$ one can rewrite this as  
\be
{1 \over 2} \Gamma^{\mu\nu} [D_{\mu},D_{\nu} ] \Gamma \cdot \Psi + \Gamma^{\nu} [D_{\mu},D_{\nu} ] 
\Psi^{\mu} - m (\slsh{D} \Gamma \cdot \Psi - D \cdot \Psi ) = 0 \,,
\ee
Using once again (\ref{RSE2}) and that the curvature tensor acts on a spinor as 
$$[D_{\mu},D_{\nu} ] \xi = {1 \over 4} R_{\mu\nu}^{~ab} (\gamma_{ab} \xi) = -  {1 \over 4} R_{\mu\nu\rho\sigma} (\Gamma^{\rho\sigma} \xi) $$
we find 
\be
R_{\mu\nu\rho\sigma} \left(- {1 \over 8} \, \Gamma^{\mu\nu} \Gamma^{\rho\sigma} \Gamma \cdot \Psi - 
{1 \over 4} \, \Gamma^{\nu} \Gamma^{\rho\sigma} \Psi^{\mu}  \right) + R_{\nu\sigma} \Gamma^{\nu} 
\Psi^{\sigma} - {d-1 \over d-2} \, m^{2} \, \Gamma \cdot \Psi = 0 \,.
\ee
After some gamma-matrix algebra, one can see that the expression involves in fact only the Ricci tensor 
$$R_{\mu\nu} = g^{\rho\sigma} R_{\mu\rho\sigma\nu} = \Lambda g_{\mu\nu} \,. $$
Finally we obtain (\ref{RSconstr}).

\section{Solution of the e.o.m. to first order}\label{solfirst}

 \noindent\({X_1=}\\
{\left\{-\frac{1}{f^{3/4} (1+f)^{1/4}}i u^{9/4} \left(-2+\sqrt{f}\right) \right.}\\
{(\text{$C_{1}$}-\, }\\
{\frac{1}{72} i \left(\frac{24 \sqrt{1+\sqrt{f}} \left(8 \qn \left(-1+\sqrt{f}\right)+3 \left(2+\sqrt{f}\right) \wn \right) \alpha _0}{-3+u^2+3
\sqrt{f}}+\right.}\\
{2 \left(18 \wn  \text{Log}[f]+32 \sqrt{3} \qn \text{Log}\left[2-\sqrt{f}\right]-72 \wn  \text{Log}\left[1+\sqrt{1+\sqrt{f}}\right]-\right.}\\
{18 \sqrt{2} \qn \text{Log}\left[\frac{\sqrt{2}-\sqrt{1+\sqrt{f}}}{\sqrt{2}+\sqrt{1+\sqrt{f}}}\right]-}\\
{27 \sqrt{2} \wn  \text{Log}\left[\frac{\sqrt{2}-\sqrt{1+\sqrt{f}}}{\sqrt{2}+\sqrt{1+\sqrt{f}}}\right]-}\\
{\left.64 \sqrt{3} \qn \text{Log}\left[\sqrt{3}+\sqrt{1+\sqrt{f}}\right]\right) \alpha _0+}\\
{64 \sqrt{3} \qn \text{Log}\left[\frac{\sqrt{3}+\sqrt{1+\sqrt{f}}}{\sqrt{3}-\sqrt{1+\sqrt{f}}}\right] \alpha _0-\frac{36 \qn \sqrt{1+\sqrt{f}} \left(\alpha
_0-i \beta _0\right)}{-1+\sqrt{f}}+}\\
{\frac{64 \qn \sqrt{1+\sqrt{f}} \left(2 \alpha _0-i \beta _0\right)}{-2+\sqrt{f}}+}\\
{\left.\left.9 \sqrt{2} \qn \text{Log}\left[\frac{\sqrt{2}-\sqrt{1+\sqrt{f}}}{\sqrt{2}+\sqrt{1+\sqrt{f}}}\right] \left(5 \alpha _0-i \beta _0\right)-\frac{8
q \left(\alpha _0+i \beta _0\right)}{\sqrt{1+\sqrt{f}}}\right)\right),}\\
{\frac{1}{f^{3/4}}u^{7/4} \left(1+\sqrt{f}\right)^{1/4} \left(2+\sqrt{f}\right) }\\
{(\text{$C_{2}$}+}\\
{\frac{1}{72} \left(\frac{216 i \sqrt{1-\sqrt{f}} \wn  \alpha _0}{1+\sqrt{f}}-\frac{96 i u (2 \qn+3 \wn ) \alpha _0}{\sqrt{1+\sqrt{f}} \left(2+\sqrt{f}\right)}+\right.}\\
{\frac{72 \sqrt{2} \left(-i+\sqrt{2}\right) \wn  \text{Log}\left[1-\sqrt{1-\sqrt{f}}\right] \alpha _0}{2 i+\sqrt{2}}-}\\
{18 i \sqrt{2} (2 \qn+3 \wn ) \left(\text{Log}\left[1+\sqrt{f}\right]-2 \text{Log}\left[\sqrt{2}-\sqrt{1-\sqrt{f}}\right]\right) \alpha _0+}\\
{72 i \wn  \text{Log}\left[1+\sqrt{1-\sqrt{f}}\right] \alpha _0+\frac{8 i \qn \left(\alpha _0+i \beta _0\right)}{\sqrt{1-\sqrt{f}}}-}\\
{\frac{36 \qn \sqrt{1-\sqrt{f}} \left(i \alpha _0+\beta _0\right)}{1+\sqrt{f}}+\frac{64 \qn u \left(2 i \alpha _0+\beta _0\right)}{\sqrt{1+\sqrt{f}}
\left(2+\sqrt{f}\right)}+}\\
{\left.\left.9 \sqrt{2} \qn \left(\text{Log}\left[1+\sqrt{f}\right]-2 \text{Log}\left[\sqrt{2}-\sqrt{1-\sqrt{f}}\right]\right) \left(5 i \alpha
_0+\beta _0\right)\right)\right),}\\
{-\frac{u^{1/4}\left(1+\sqrt{f}\right)^{1/4}}{72 f^{1/4}}}\\
{\left(72 \sqrt{1+\sqrt{f}} \left(\text{$C_{3}$}-i \text{$C_{1}$} \sqrt{f}\right)+\right.}\\
{\left(9 \sqrt{2} \sqrt{1+\sqrt{f}} (q-6 \wn ) \text{Log}\left[\sqrt{2}-\sqrt{1+\sqrt{f}}\right]-\right.}\\
{3 \left(92 \qn+24 \wn +3 \sqrt{2} \sqrt{1+\sqrt{f}} (q-6 \wn ) \text{Log}\left[\sqrt{2}+\sqrt{1+\sqrt{f}}\right]\right)+}\\
{\sqrt{f} \left(-180 \qn+72 \wn -36 \sqrt{1+\sqrt{f}} \wn  \text{Log}[f]-\right.}\\
{64 \sqrt{3} \qn \sqrt{1+\sqrt{f}} \text{Log}\left[2-\sqrt{f}\right]-}\\
{9 \sqrt{2} \qn \sqrt{1+\sqrt{f}} \text{Log}\left[\sqrt{2}-\sqrt{1+\sqrt{f}}\right]+}\\
{54 \sqrt{2} \sqrt{1+\sqrt{f}} \wn  \text{Log}\left[\sqrt{2}-\sqrt{1+\sqrt{f}}\right]+}\\
{64 \sqrt{3} \qn \sqrt{1+\sqrt{f}} \text{Log}\left[\sqrt{3}-\sqrt{1+\sqrt{f}}\right]+}\\
{144 \sqrt{1+\sqrt{f}} \wn  \text{Log}\left[1+\sqrt{1+\sqrt{f}}\right]+}\\
{9 \sqrt{2} \qn \sqrt{1+\sqrt{f}} \text{Log}\left[\sqrt{2}+\sqrt{1+\sqrt{f}}\right]-}\\
{54 \sqrt{2} \sqrt{1+\sqrt{f}} \wn  \text{Log}\left[\sqrt{2}+\sqrt{1+\sqrt{f}}\right]+}\\
{\left.\left.64 \sqrt{3} \qn \sqrt{1+\sqrt{f}} \text{Log}\left[\sqrt{3}+\sqrt{1+\sqrt{f}}\right]\right)\right) \alpha _0+}\\
{3 i \left(-20 \qn+12 \qn \sqrt{f}-48 \wn +3 \sqrt{2} \qn \left(-1+\sqrt{f}\right) \sqrt{1+\sqrt{f}} \right.}\\
{\text{Log}\left[\sqrt{2}-\sqrt{1+\sqrt{f}}\right]-24 \sqrt{1+\sqrt{f}} \wn  \text{Log}\left[-1+\sqrt{1+\sqrt{f}}\right]+}\\
{24 \sqrt{1+\sqrt{f}} \wn  \text{Log}\left[1+\sqrt{1+\sqrt{f}}\right]+}\\
{3 \sqrt{2} \qn \sqrt{1+\sqrt{f}} \text{Log}\left[\sqrt{2}+\sqrt{1+\sqrt{f}}\right]-}\\
{\left.\left.3 \sqrt{2} \qn \sqrt{f \left(1+\sqrt{f}\right)} \text{Log}\left[\sqrt{2}+\sqrt{1+\sqrt{f}}\right]\right) \beta _0\right),}\\
{\frac{1}{18 f^{1/4} \left(1+\sqrt{f}\right)^{3/4}}i u^{7/4} }\\
{(18 \text{$C_{4}$}+}\\
{i \left(18 \text{$C_{2}$} \sqrt{f}-18 i \sqrt{f} \wn  \text{Log}\left[1-\sqrt{1-\sqrt{f}}\right] \alpha _0+\right.}\\
{18 i \sqrt{f} \wn  \text{Log}\left[1+\sqrt{1-\sqrt{f}}\right] \alpha _0-}\\
{\frac{9 \left(-i+\sqrt{2}\right) \sqrt{f} \text{Log}\left[1+\sqrt{f}\right] \left((q-6 \wn ) \alpha _0-i \qn \beta _0\right)}{2 \left(2 i+\sqrt{2}\right)}+}\\
{\frac{1}{2 i+\sqrt{2}}9 \left(-i+\sqrt{2}\right) \sqrt{f} \text{Log}\left[\sqrt{2}-\sqrt{1-\sqrt{f}}\right] }\\
{\left((q-6 \wn ) \alpha _0-i \qn \beta _0\right)-i \sqrt{1-\sqrt{f}} \left((31 \qn-126 \wn ) \alpha _0-23 i \qn \beta _0\right)+}\\
{\frac{45}{2
\sqrt{2}} \text{Log}\left[\frac{\sqrt{2}+\sqrt{1-\sqrt{f}}}{\sqrt{2}-\sqrt{1-\sqrt{f}}}\right] \left(i (q-6 \wn ) \alpha _0+q \beta _0\right)+}\\
{\frac{1}{\sqrt{1-\sqrt{f}}}}\\
{4 \left(-i \left(-25 \qn+\sqrt{f} (19 \qn-36 \wn )+27 \wn \right) \alpha _0+\right.}\\
{\left.\left(q \left(2-8 \sqrt{f}\right)-9 \wn \right) \beta _0\right)+}\\
{18 \left(-i \sqrt{2} (q-6 \wn ) \text{Log}\left[\frac{\sqrt{2}+\sqrt{1-\sqrt{f}}}{\sqrt{1+\sqrt{f}}}\right] \alpha _0+\right.}\\
{\left.\left.\left.\left.\left(\wn  \text{Log}\left[\frac{1+\sqrt{1-\sqrt{f}}}{1-\sqrt{1-\sqrt{f}}}\right]-\sqrt{2} \qn \text{Log}\left[\frac{\sqrt{2}+\sqrt{1-\sqrt{f}}}{\sqrt{1+\sqrt{f}}}\right]\right)
\beta _0\right)\right)\right)\right\};}\)


\newpage

\begin{thebibliography}{99}

\bibitem{Policastro:2001yc}
  G.~Policastro, D.~T.~Son and A.~O.~Starinets,
  Phys.\ Rev.\ Lett.\  {\bf 87} (2001) 081601
  [arXiv:hep-th/0104066].

\bibitem{Policastro:2002se}
  G.~Policastro, D.~T.~Son and A.~O.~Starinets,
  JHEP {\bf 0209} (2002) 043
  [arXiv:hep-th/0205052].

\bibitem{Policastro:2002tn}
  G.~Policastro, D.~T.~Son and A.~O.~Starinets,
  JHEP {\bf 0212} (2002) 054
  [arXiv:hep-th/0210220].
  
  \bibitem{landau5}
  L.D.~Landau and E.M.~Lifshitz, {\it Fluid mechanics}, Pergamon Press, NY, 1987. 

\bibitem{Aharony:1999ti}
  O.~Aharony, S.~S.~Gubser, J.~M.~Maldacena, H.~Ooguri and Y.~Oz,
  Phys.\ Rept.\  {\bf 323} (2000) 183
  [arXiv:hep-th/9905111].


\bibitem{Kaminski:2008ai}
  M.~Kaminski,
  arXiv:0808.1114 [hep-th].

\bibitem{Lebedev:1989rz}
  V.~V.~Lebedev and A.~V.~Smilga,
  Nucl.\ Phys.\  B {\bf 318} (1989) 669.

\bibitem{Kovtun:2003vj}
  P.~Kovtun and L.~G.~Yaffe,
  Phys.\ Rev.\  D {\bf 68} (2003) 025007
  [arXiv:hep-th/0303010].
  
  

%
\bibitem{Corley:1998qg}
  S.~Corley,
  Phys.\ Rev.\  D {\bf 59} (1999) 086003
  [arXiv:hep-th/9808184].


\bibitem{Volovich:1998tj}
  A.~Volovich,
  JHEP {\bf 9809} (1998) 022
  [arXiv:hep-th/9809009].


\bibitem{Rashkov:1999ji}
  R.~C.~Rashkov,
  Mod.\ Phys.\ Lett.\  A {\bf 14} (1999) 1783
  [arXiv:hep-th/9904098].

\bibitem{Henningson:1998cd}
  M.~Henningson and K.~Sfetsos,
  Phys.\ Lett.\  B {\bf 431} (1998) 63
  [arXiv:hep-th/9803251].

\bibitem{Grassi:2000dm}
  P.~A.~Grassi and P.~van Nieuwenhuizen,
``No van Dam-Veltman-Zakharov discontinuity for supergravity in AdS  space,''
 Phys.\ Lett.\  B {\bf 499} (2001) 174
 [arXiv:hep-th/0011278].

\bibitem{Skenderis:2002wp}
  K.~Skenderis,
  Class.\ Quant.\ Grav.\  {\bf 19} (2002) 5849
  [arXiv:hep-th/0209067].


\bibitem{Bhattacharyya:2008jc}
  S.~Bhattacharyya, V.~E.~Hubeny, S.~Minwalla and M.~Rangamani,
  JHEP {\bf 0802} (2008) 045
  [arXiv:0712.2456 [hep-th]].
  
\bibitem{Iqbal:2008by}
  N.~Iqbal and H.~Liu,
  arXiv:0809.3808 [hep-th].
  
 
\end{thebibliography}
\end{document}